\documentclass[twocol]{ametsoc}

\journal{jas}

\bibpunct{(}{)}{;}{a}{}{,}

\title{Inertial waves in axisymmetric tropical cyclones}

    \authors{Morgan E O'Neill\correspondingauthor{Morgan E O'Neill, 
     Dept. of Earth System Science, 
     Stanford University, Stanford, CA 94305.}}

     \affiliation{Dept. of Earth System Science, Stanford University, Stanford, CA, USA}

\email{morgan.e.oneill@gmail.com}

    \extraauthor{Daniel R. Chavas}
    \extraaffil{Dept. of Earth, Atmospheric, and Planetary Sciences, Purdue University, West Lafayette, IN, USA}

\abstract{The heat engine model of tropical cyclones describes a thermally direct overturning circulation. Outflowing air slowly subsides as radiative cooling to space balances adiabatic warming, a process that does not consume any work. However, we show here that the lateral spread of the outflow is limited by the environmental deformation radius, which at high latitudes can be rather small. In such cases, the outflowing air is radially constrained, which limits how far downward it can subside via radiative cooling alone. Some literature has invoked the possibility of `mechanical subsidence' or `forced descent’ in the storm outflow region in the presence of high inertial stability, which would be a thermally indirect circulation. Mechanical subsidence in the subsiding branch of a tropical cyclone has not before been observed or characterized. A series of axisymmetric tropical cyclone simulations at different latitudes and domain sizes is conducted to study the impact of environmental inertial stability on storm dynamics. In higher latitude storms in large axisymmetric domains, the outflow acts as a wavemaker to excite an inertial wave at the environmental inertial (Coriolis) frequency. This inertial wave periodically ventilates the core of a high-latitude storm with its own low entropy exhaust air. The wave response is in contrast to the presumed forced descent model, and we hypothesize that this is because inertial stability provides less resistance than buoyant stability, even in highly inertially stable environments.} 

\begin{document}

\maketitle

%
\section{Introduction}




There remains a large gap in understanding concerning the dynamics and downstream impacts of the TC outflow. \citet{SchenkelHart2015} showed that a TC's impact on the tropics after passage can be surprisingly intense and enduring: large negative moist static energy signals in the wake of Pacific typhoons lasted beyond 40 days after passage. A better understanding of how the TC outflow alters the larger-scale tropospheric environment may also help us constrain the upper bound on global annual TC frequency \citep{Hoogewindetal2019}, and how far the real Earth system is from that upper bound.


Outflow anticyclones in the upper troposphere, responsible for transporting high entropy, low angular momentum air away from the TC core, can extend well beyond 2000 km from the TC center in vast, highly asymmetric outflow jets \citep{Merrill1988a,MerrillVelden1996,Ditcheketal2017}. \citet{Merrill1988a} noted that asymmetric TC outflow structures have typical scales of 3000 km in the upper troposphere. These massive outflow jets clearly demonstrate that the TC secondary circulation is not closed - much of that air is never going to recycle through the TC storm a second time. TCs instead constantly ingest `new' air while some processed air is advected far away (most likely for fast-moving storms or those near midlatitude jets). For example, TCs visibly ingest dry air from the Saharan Air Layer \citep{DunionVelden2004}. High-enthalpy air in this case leaves the TC via the outflow and is replaced by low-enthalpy air because the secondary circulation is not closed. But what is that fraction? How much air is processed just once by the TC, and where does it go afterward? What controls where the outflow air goes and how it behaves? 

In the theoretical Carnot heat engine model of TCs \citep{Emanuel1986,Emanuel1988b,BisterEmanuel1998}, the secondary circulation is closed and thermally direct, creating work that drives the most visible and important aspects of a TC (recently updated to not require a closed circulation by \citet{RousseauRizziEmanuel2019}). Most work is expended to spin up and maintain cyclonic winds against friction, loft water into the atmosphere \citep{PauluisZhang2017}, and, during the growth phase, expand the outflow against the surrounding environment \citep{Rappinetal2011}. Among the scales of thermally direct convection in Earth's atmosphere, the hurricane heat engine may be the most efficient, given \citet{PauluisZhang2017}'s recent estimate that air parcels in the eyewall experienced 70\% of the efficiency of a perfect Carnot engine in a three-dimensional numerical simulation. A developing TC does work on its environment as it expands the outflow; however a mature TC does no useful work on its environment \citep{Bisteretal2010}.

For a mature, steady-state TC in this idealized framework, the outflow air is assumed to exhaust at large radii. This provides sufficient time for radiative cooling to be balanced by adiabatic warming as each parcel sinks back down to the boundary layer, a process that assumes that the environment does not provide resistance to the subsiding outflow air. Carnot theory does not consider resistance from inertial stability of the environment.

In the real atmosphere, environmental conditions can provide substantial resistance to the divergent TC outflow. Inertial stability, which is a measure of resistance to symmetric radial motions in a spinning fluid, varies greatly in the subtropics where many TCs propagate, and can be large. In axisymmetric flow, inertial stability is 
\begin{equation}
I^2=\frac{1}{r^3}\frac{\partial M^2}{\partial r}
\end{equation}
where $M=rv+1/2f r^2$ is the absolute angular momentum for tangential wind $v$ and radius $r$. On an $f$-plane in solid body rotation, environmental inertial stability $I^2$ would simply be a function of latitude $\phi$: $I^2=f^2=(2\Omega \sin\phi)^2$. With horizontal flow that experiences azimuthal asymmetries, angular momentum is not materially conserved and $I^2$ is more complicated.

On Earth, inertial stability and the corresponding resistance to horizontal divergence typically increase away from the equator, but are strongly modulated by features like jets, troughs and tropical phenomena. Studies of TC outflow jets argue that TC exhaust air preferentially flows away from TCs in highly asymmetric jets \citep{BlackAnthes1971,Merrill1988a,Rappinetal2011} toward where inertial resistance, or $I^2$, is minimized, thus requiring the least work. 

\citet{Rappinetal2011} carried out a numerical study using three-dimensional simulations of tropical cyclones to understand the role of environmental inertial stability in cyclogenesis and intensification rates of TCs. They showed that storms in low inertial stability environments were able to expand an anticyclone against the environment more quickly (and thus intensify more rapidly), because the work required to displace the environment was reduced. However, implicit in \citet{Rappinetal2011}'s discussion is an assumption that a TC expends work to force descent of air in the outflow region under high inertial stability: ``...outflow expands out to the Rossby radius, where the outflow ``searches'' for a weakness in the inertial wall. If one is found, and the work required for further expansion is less than that required to force subsidence at the Rossby radius, the outflow is ventilated through that weakness so that further radiational cooling reduces the energy drain of forced subsidence against buoyancy.'' However, this forced subsidence has never been characterized or quantified, in contrast to a broad literature on the thermally indirect nature of the TC eye (e.g., \citet{Willoughby1979,Smith1980,Emanuel1997,Willoughby1998,ZhangKieu2006}). 

We first test the hypothesis that the outflow radius scales with the deformation radius, as suggested by \citet{Rappinetal2011}. We then quantify the role of environmental inertial stability on the subsiding branch of the TC overturning circulation, specifically looking at whether and how it interacts with the inner region of the TC, using a suite of axisymmetric TC simulations. We find that the theoretical assumption that the subsiding branch does not consume work may be poor at higher latitudes. However `mechanical subsidence' also appears to be an inaccurate description of what occurs when environmental inertial stability is high, because it indicates that work is consumed by pushing buoyant air downward. In contrast, our results show that subsiding air instead makes large radial excursions back toward the center of the storm as one phase of an inertial wave that propagates upward from the boundary layer. The wavemaker is the outflow jet itself, which unsteadily impinges on the inertially stable environment. This makes sense as the natural response to high $I^2$ because even at high latitudes, resistance to vertical motion in the far environment is still much larger than resistance to horizontal motion: the squared Brunt-Vaisala frequency $N^2 >> I^2$.

The remainder is organized as follows: Section \ref{model} describes the experiments conducted using the axisymmetric cloud resolving Bryan Cloud Model 1 (CM1) \citep{BryanFritsch2002,BryanRotunno2009}, and Section \ref{evolution} provides a discussion of storm evolution and character for each experiment. Section \ref{streamfunctions} reviews the time-mean streamfunctions of the simulated TCs. Section \ref{wavesec} provides the results of the large-domain simulations with emphasis on inertial wave behavior at higher latitudes. Section \ref{conclusion} concludes with a discussion.

\section{Experimental setup}\label{model}

Using the Bryan Cloud Model (CM1) \citep{BryanFritsch2002} version 19.5 in axisymmetric geometry, we run two different sets of simulations. Within a set, 13 simulations are run, each at a constant latitude between 10$^\circ$N to 40$^\circ$N (same range as \cite{DeMariaPickle1988a}) in regular intervals of 2.5$^\circ$. As small-domain simulations are common in the literature, and known to artificially limit storm size, we seek to examine how results differ when the domain size is smaller than the environmental deformation radius. One simulation set has a domain radial extent of 1500 km ($R_\textrm{small}$) and the other set has an extent of 6000 km ($R_\textrm{large}$). Every simulation employs a lateral sponge layer in the outermost 100 km, which restores the environment to its background state, acting as a source or sink of any angular momentum that deviates from solid-body rotation and preventing waves from reflecting back into the domain.

The version of CM1 employed here solves the compressible, nonhydrostatic, axisymmetric Reynolds-averaged (\texttt{cm1setup=2}) equations using a Klemp-Wilhelmson \citep{KlempWilhelmson1978} time stepping method (using CM1 option \texttt{psolver=3} in the \texttt{namelist.input} file) similar to that used in the Weather Research and Forecasting Model (WRF; \cite{Skamarocketal2005}), as described in \citet{BryanRotunno2009}.
The domain is 25 km high with a sponge layer in the uppermost 2 km, with a stretched vertical resolution (using cm1 option \texttt{stretch}\_\texttt{z}$=$\texttt{1} that employs a stretching algorithm from \citet{WilhelmsonChen1982}). There are 61 vertical gridpoints that preferentially resolve the boundary layer, with a lowest vertical resolution of 50 m that is smoothly stretched to reach a constant 0.5 km vertical resolution above 5.25 km. The horizontal resolution is constant at 4 km. The axisymmetric prognostic equations solved by the model are similar to those in \citep{BryanRotunno2009}, using what they refer to as the `traditional equation set' for moist microphysics (\texttt{eqtset=1}), though updated to include water ice variables as described at  {\url{http://www2.mmm.ucar.edu/people/bryan/cm1/cm1_equations.pdf}}, last modified on September  5, 2017. Thompson microphysics \citep{Thompsonetal2008} is used and the model includes dissipative heating. The horizontal mixing length $l_h$ is a linear function of surface pressure defined by the points $l_h(p_s=1015\textrm{hPa})=100$ m and $l_h(p_s=900 \textrm{hPa})= 1000$ m. A Smagorinsky type horizontal turbulence closure is used \citep{Smagorinsky1963}. That scheme as well as the planetary boundary layer parameterization scheme are described in \citet{BryanRotunno2009}. Output was saved as a snapshot every two hours.

Insolation occurs on the diurnal cycle for a constant equivalent latitude of $20^{\circ}$N on May 15th for all simulations and the interactive radiative scheme is the Rapid Radiative Transfer Model for global climate models (RRTMG as adapted from the WRF model). The model is initialized with the \citet{RotunnoEmanuel1987} vortex and the initial environmental state is set by the \citet{Dunion2011} moist tropical sounding. SST is fixed at 301K. All simulations are run for 200 days, to give each storm an opportunity to reach a quasi-equilibrium state for at least several tens of days (e.g. \citet{ChavasEmanuel2014}).

\section{Storm evolution and size}\label{evolution}

\begin{figure*}[htbp]
\centerline{\includegraphics[width=39pc]{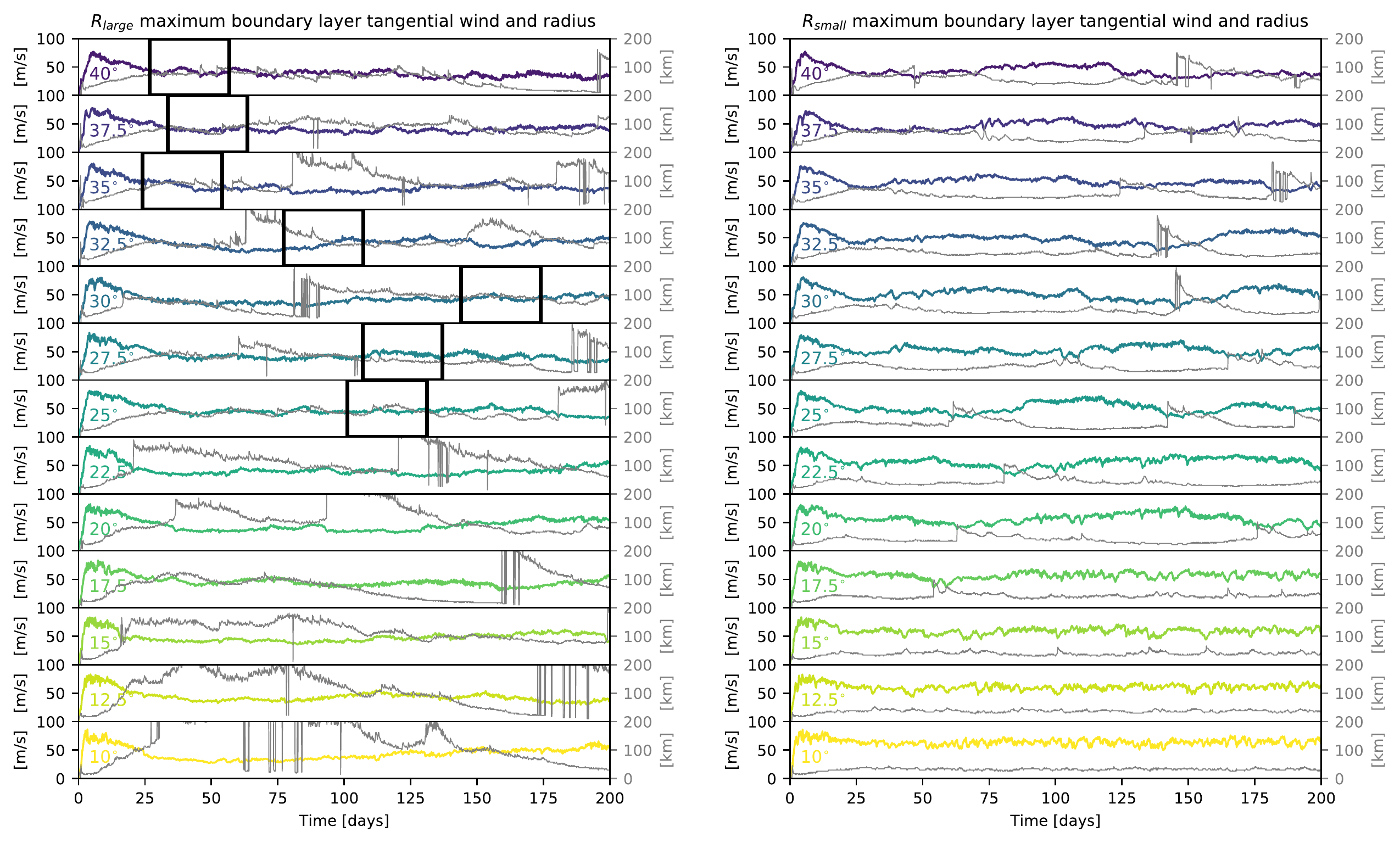}}
\caption{$v_{max}$ (m/s, colors) and $r_{max}$ (km, grey) at the lowest model level as a function of day since start of simulation for simulations $R_\textrm{large}$ (left) and $R_\textrm{small}$ (right). The selected time windows in boxes are examined in later sections because they experience the strongest inertial oscillation.}\label{windtimeRR}
\end{figure*}

Every experiment develops a strong tropical cyclone by day five (Fig. \ref{windtimeRR}), with maximum surface wind speeds approaching $v_{max} = 100 \; ms^{-1}$ among lower-latitude storms. $v_{max}$ subsequently decreases to long-term values ranging between 40-70$\; ms^{-1}$ for the remainder of the simulations. The radius of maximum wind (RMW) of low latitude storms is significantly smaller in $R_\textrm{small}$ simulations compared to $R_\textrm{large}$ simulations. All $R_{large}$ simulations exhibit a much larger RMW than observed on Earth for a simulated equilibrium intensity of approximately $50 \; ms^{-1}$ \citep{Sternetal2015}. Both simulated characteristics were also observed in the simulations of \citet{ChavasEmanuel2014}.

Nearly every storm experiences at least one large structural disruption to its RMW and $v_{max}$ over the integration period. This event involves an outer cyclonic `jet' that propagates inward to eventually replace the primary eyewall. The characteristic sign of this event is a decrease in $v_{max}$ until the RMW suddenly jumps to a larger radius, where the incoming jet and its peak cyclonic wind become identified as the new eyewall. This is qualitatively similar to an eyewall replacement cycle (e.g. \citet{Sitkowskietal2011}), and has been observed in axisymmetric models in \citet{Hakim2011,Hakim2013,ChavasEmanuel2014,Frisius2015}. The only exceptions -- storm that don't appear to experience an ERC-like event -- are $R_\textrm{small}$($10^\circ$) and potentially $R_\textrm{small}$($12.5^\circ$). These are the largest storms in the smallest domain, minimizing or precluding any environmental variability independent of the TC circulation itself. An earlier set of simulations using RRTM radiation with a domain extent to 8000 km, run out to 300 days, demonstrated that massive ERC-like events occurred at every latitude (not shown).

These ERC-like events, which occur at very large radii and over much slower timescales, are not observed in nature but instead are an artifact of axisymmetric geometry. Every parcel in an axisymmetric model is actually a ring of air, and therefore must conserve angular momentum during radial motion. Additionally, moist convection is free to occur throughout the domain. The consequence of these two choices in concert is that in $R_\textrm{large}$ storms, deep convection occurs well away from the TC, as observed in \citet{Frisius2015,Persingetal2019}, and persists for many tens of days. Conservation of angular momentum causes each convective tower to establish counter-rotating `jets' at the lower levels where air converges and upper levels where air diverges, leading to concentric convective rings not unlike nested eyewalls. In a real three dimensional environment, a convective tower far away from the TC center is  highly localized and does not materially conserve angular momentum, and eddy angular momentum flux convergence plays a major role in the outflow (e.g., \citet{Anthes1974,Ditcheketal2017}). The present simulations thus exhibit some unrealistic behavior including occasional secondary, elevated tangential wind maxima within massive eyes, ERC-like events occurring beyond 200 km radially, and eyewalls that are slanted nearly horizontally.

\begin{figure}[h]
\centerline{\includegraphics[width=19pc]{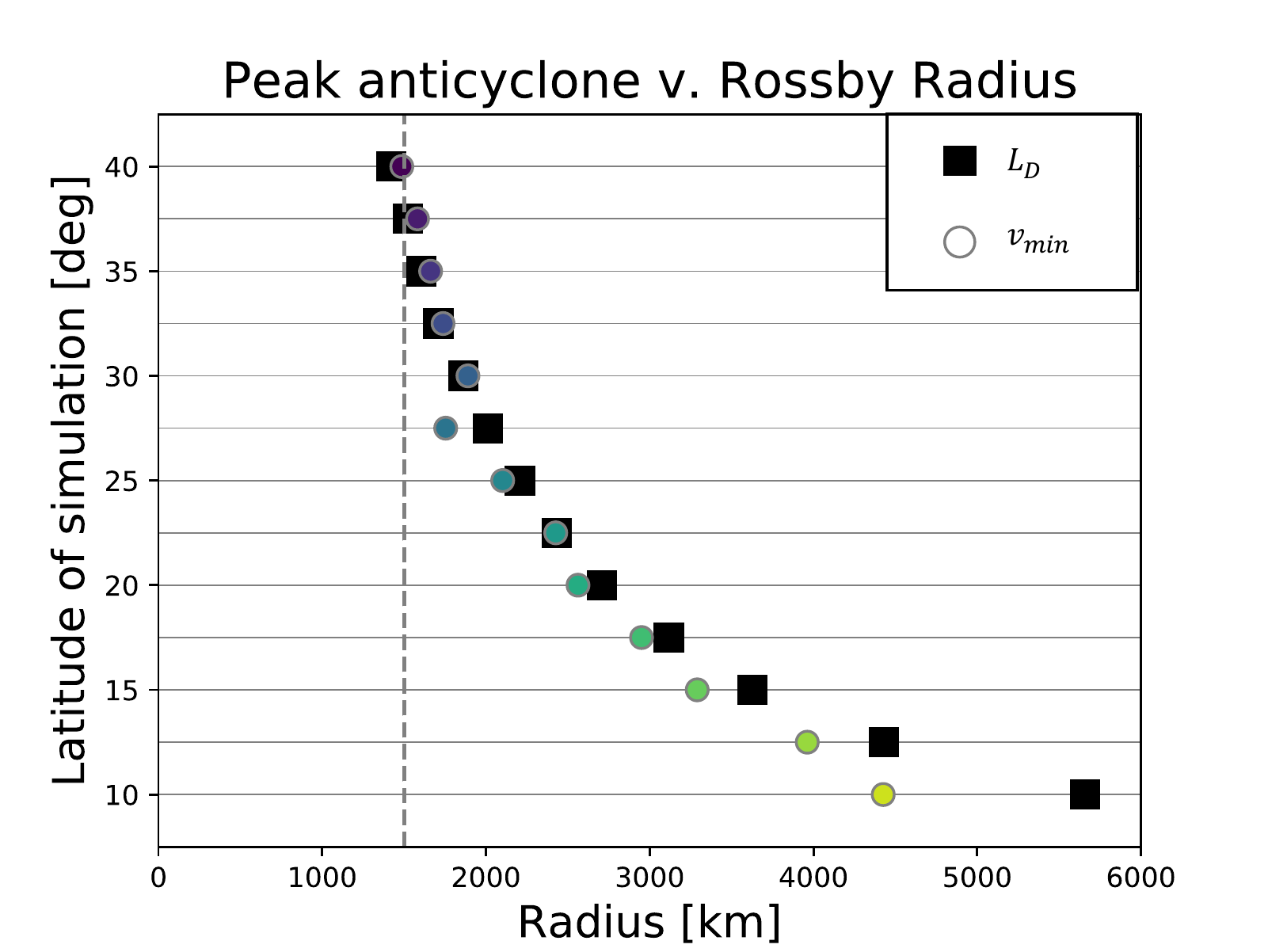}}
\caption{Location of the core of the anticyclonic jet (minimum tangential wind $v_{min}$, circles) and deformation radius (squares) as a function of latitude. The grey dashed line marks the domain size of experiments $R_\textrm{small}$.}\label{Ldoutflow}
\end{figure}

One expectation as mentioned in \citet{Rappinetal2011} is robustly met across all $R_\textrm{large}$ storms - the outflow anticyclone location scales well with the environmental deformation radius $L_D = NH/f$ (Fig. \ref{Ldoutflow}), where $H$ is the depth of the free troposphere defined as the vertical distance between the model-output planetary boundary layer height (varying between 1.7 and 2 km) and the tropopause (13.75 km). The squared buoyancy frequency is $N^2=g/\theta_v \partial \theta_v/\partial z$ for virtual potential temperature $\theta_v =\theta (1+q_v/\epsilon)/(1+q_v)$, dry potential temperature $\theta$, water vapor mixing ratio $q_v$, and $\epsilon=R_d/R_v$ is the ratio of the dry air and water vapor gas constants. $L_D$ was calculated using an environmental sounding averaged over the outermost 100 km of the free troposphere (excluding/radially inward of the horizontal sponge layer) from day 25 to 200 for each simulation. The outflow anticyclone location (circle) is marked at the radial location of the maximum anticyclonic winds. Such a close fit to $L_D$ demonstrates the enormity of idealized TCs if given sufficient space and time in a numerical simulation. In contrast the $R_\textrm{small}$ simulations always have a peak anticyclonic jet at or just shy of 1400 km where the sponge layer of the outer boundary starts (not shown), and the environmental deformation radius can't be measured because there is no environment apart from the storm circulation itself.

The expansion of the outflow to the deformation radius was stated but not demonstrated in the 3D study of \cite{Rappinetal2011}. 
Figure \ref{Ldoutflow} is the first explicit numerical demonstration that axisymmetric TC outflow expands to the deformation radius in steady state. \citet{ChavasEmanuel2014} found, using a similar setup of CM1 in axisymmetry, that in contrast to the outflow $L_D$ scaling found here, the size of the surface circulation scales with $v_p/f$ for a potential intensity $v_p$. Thus the sizes of the outflow and surface circulations may scale differently.

\section{The secondary circulation}\label{streamfunctions}

A few papers have shown the full overturning circulation, including the downwelling branch, in modern axisymmetric models: \citet{EmanuelRotunno2011} (averaged over 24 hours) and \citet{Allandetal2017} (averaged over 12 hours). To our knowledge, only \citet{Frisius2015} published a streamfunction of an axisymmetric TC simulation averaged over a long time period (in that work, from 105 to 120 days). His simulations are remarkably steady in time after TC development, and this is likely due to an additional moisture sink term to dampen environmental variability. For the present experiments with large domains out to 6000 km and sophisticated radiation and microphysics, we find that the mass streamfunctions can hide behavior of a storm that can undergo intermittent periods of dramatic variability due to interaction with its environment. If the period over which an average is taken includes an ERC-like event, the upward eyewall branch appears extremely wide with low vertical velocities. A very long time average over almost any of these simulations would be affected by ERC-like events to varying degrees. First we will discuss the long-term average and then identify and study more well-behaved 30 day windows during the TC lifetime.

\begin{figure*}[h]
\centerline{\includegraphics[width=39pc]{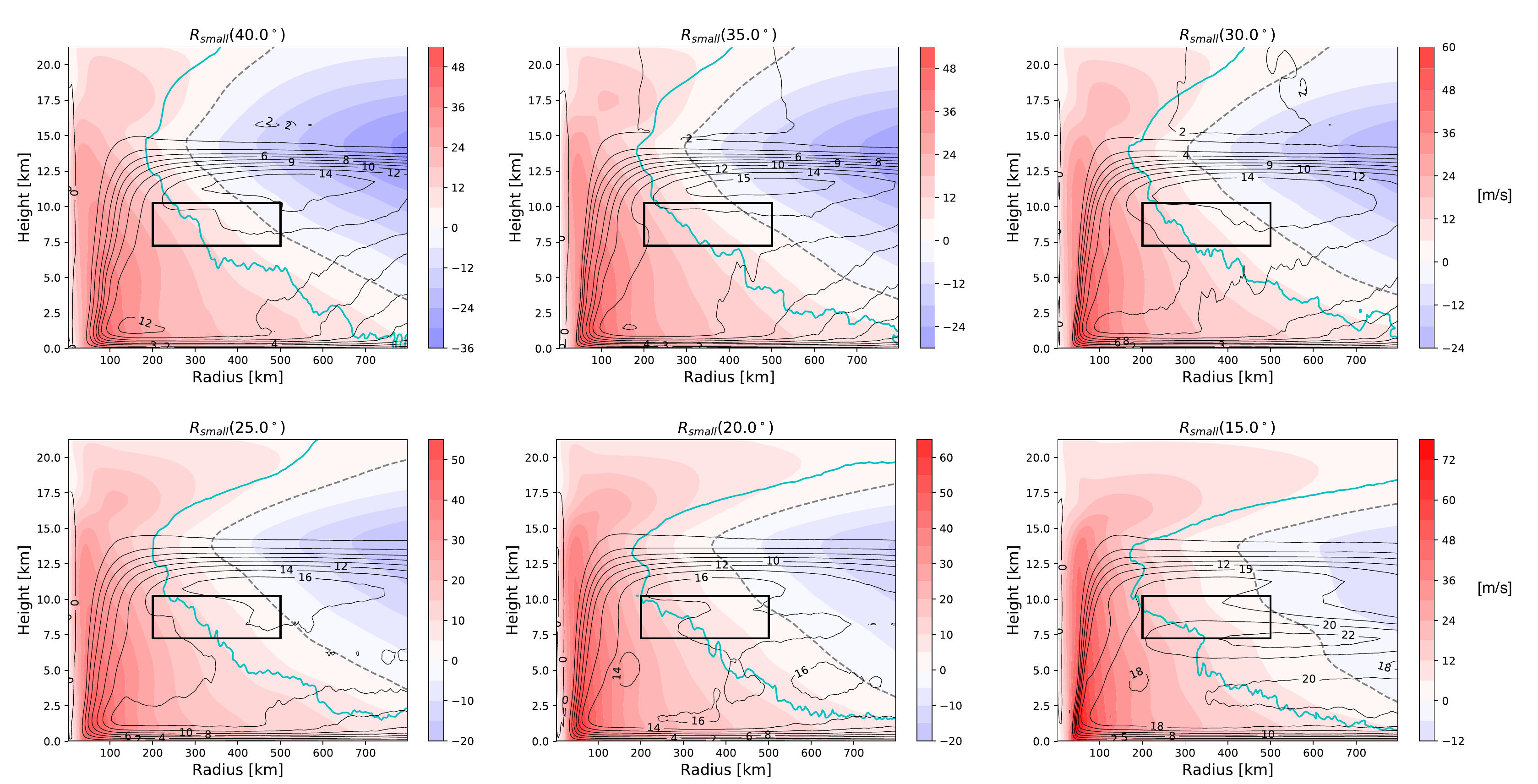}}
\caption{Tangential wind ([m/s]; colors; zero is dashed grey) and mass streamfunction [$10^8$ kg s$^{-1}$] for selected $R_\textrm{small}$ storms averaged from day 25 to day 200. The black rectangle indicates the region used for mass-weighted time series and wave analysis. The cyan contour indicates where $I^2=f^2$.}\label{sgstreams_longavg}
\end{figure*}

\begin{figure*}[h]
\centerline{\includegraphics[width=39pc]{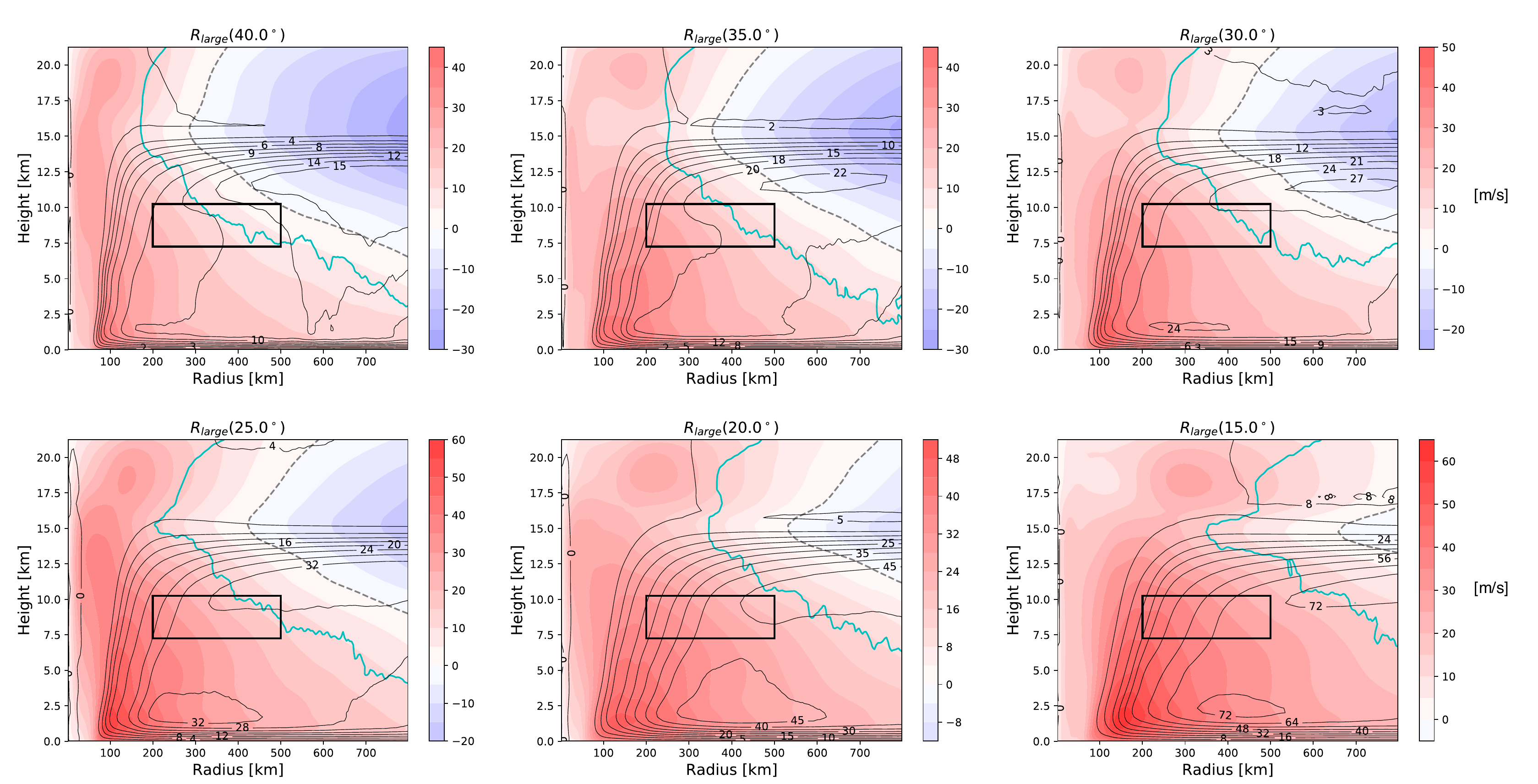}}
\caption{As Fig. \ref{sgstreams_longavg}, for days 25-200 of selected $R_\textrm{large}$ storms.} \label{gstreams_longavg}
\end{figure*}

The impact of ERCs can be seen in Figs. \ref{sgstreams_longavg} and \ref{gstreams_longavg}, which show 25-200 day averages of $R_\textrm{small}$ and $R_\textrm{large}$ storms, respectively. Intervals of $5^\circ$ from $15^\circ$ to $40^\circ$ are shown. Fig. \ref{gstreams_longavg} in particular exhibits the impact of disruptive ERC-like events on the width of the eyewall over a long time average. $R_{small}$ storms have much smaller eyes and RMW than $R_{large}$ storms at latitudes. Multiple storms exhibit a weak secondary tangential wind maximum in the stratosphere, likely due to the combination of a long-term stationary storm and realistic radiation as the primary (and slow) means of damping stratospheric circulations. However these upper level maxima appear not to participate meaningfully in the secondary circulation, which is the interest of this paper. 

Each plot has a rectangle in it that denotes an annulus; in Section \ref{wavesec}, we calculate a mass-weighted average of a number of variables within this region to examine one-dimensional time series. It extends from 7.25 km to 10.25 km in height, and from 200 km to 500 km radially -- we will refer to this region as ``the box''. It lies under the outflow of most storms, in a region commonly depicted as experiencing inward radial flow \citep{EmanuelRotunno2011,Frisius2015,Allandetal2017}. The box outer bound is 500 km, the outermost radius at which the axisymmetric assumption of a real TC circulation is tolerable \citep{Anthes1974}, because we want to study features that may have relevance to real three dimensional TCs. Indeed, \citet{RuppertONeill2019} found that a similar region experiences a robust reversal in radial wind over a diurnal cycle in a three-dimensional simulation of a TC at 20$^\circ$. To what extent is that result present or modified for axisymmetric TCs, and does environmental inertial stability matter? The box is a simple proxy for the complex changes in the overturning circulation across all of the simulations.


The mean mass streamfunctions (black labeled lines, units of [$10^8$ kg/s]) vary due to both domain size and latitude. Large domains obviously allow, and consistently yield, wider storms (e.g., \citet{Hakim2011,ChavasEmanuel2014}). The eyewall location is always larger in $R_\textrm{large}$ TCs than $R_\textrm{small}$ TCs, with broader eyes sloping outward substantially less than $45^{\circ}$ from the horizontal (also noted by \citet{Persingetal2019}). The fraction of total air that subsides through the troposphere back toward the boundary layer within 800 km radially is much smaller in $R_\textrm{large}$ TCs as well, because there is so much more room available for subsidence. The mass streamfunction illustrates the impact that domain size has on how much mass a TC overturns: $R_{small}$ mass streamfunction maxima don't vary by more than 50\%  as a function of latitude. Instead of varying substantially in size, which is precluded by the small domain, $R_{small}(10^\circ)$ storms vary substantially in intensity with the most intense storms at the lowest latitudes. However this too is not easy to interpret directly from long-term streamfunctions, and is partly due to substantial ERC-events which smear out the location and intensity of the strongest winds. $R_{large}$ TCs on the other hand experience a five-fold decrease in secondary circulation strength (streamfunction maximum) from 15$^\circ$ to 40$^\circ$. The size and structure of $R_{small}$ and $R_{large}$ appear to converge around 37.5$^\circ$-40$^\circ$ (Figs. \ref{sgstreams_longavg}a and \ref{gstreams_longavg}a), as storm size is no longer strongly influenced by the domain size even for the small domain.

$R_\textrm{small}$ storms at low latitudes exhibit a vertical standing wave in the outer streamfunction in steady state, with a vertical wavenumber of three or so within the troposphere. These are shallow, weak, stacked overturning circulations superimposed on the main overturning circulation at large radii. This odd behavior appears mostly avoided once the latitude is 22.5$^{\circ}$ or higher, and don't appear to participate in the boundary layer or inner core flow.


\section{The inertial wave}\label{wavesec}

\begin{figure*}[h]
\centerline{\includegraphics[width=29pc]{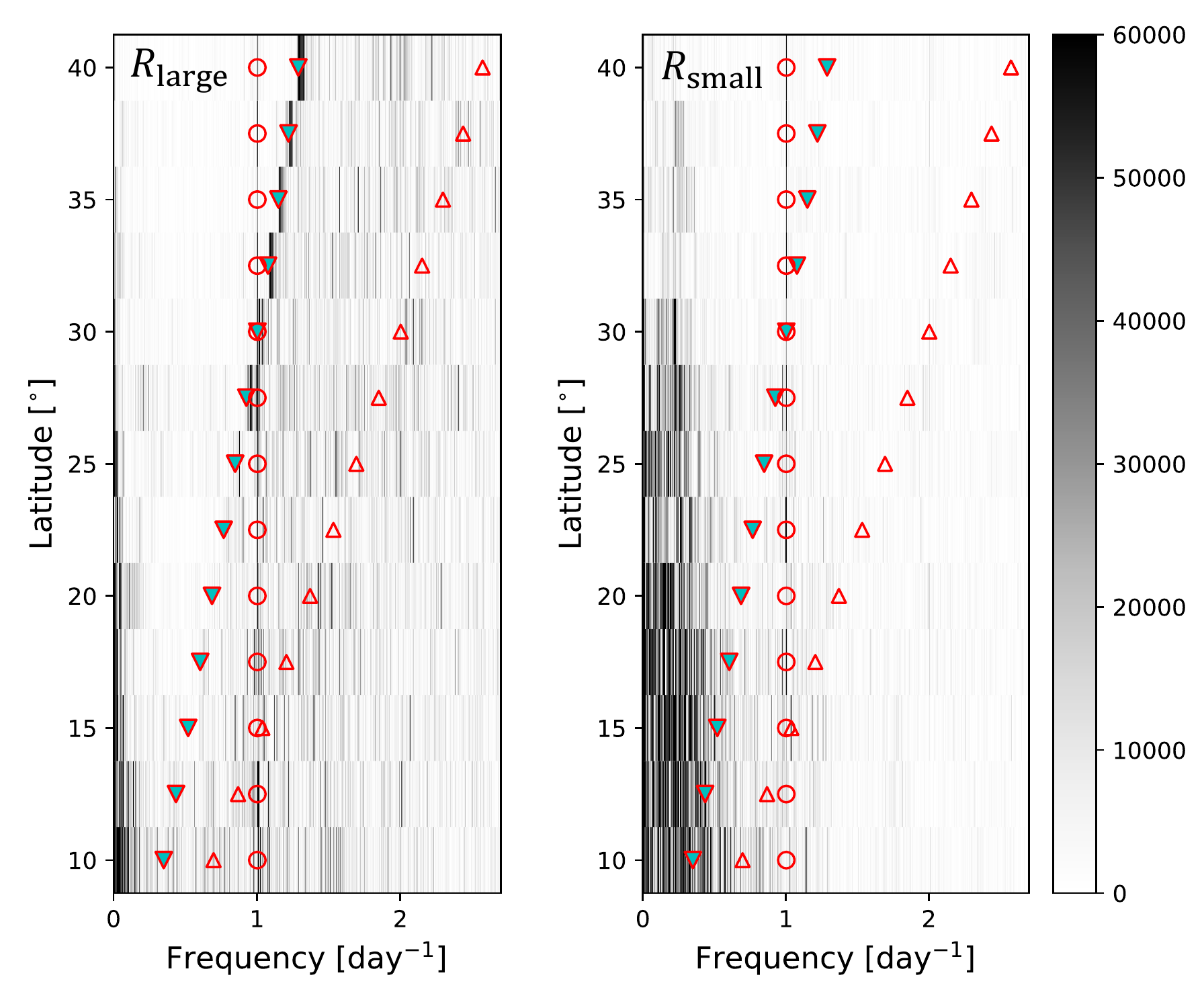}}
\caption{Power spectral density [W/Hz] of radial wind from day 15 to 200. Filled triangles indicate the Coriolis frequency $f$, open triangles indicate $2f$ and open circles indicate the diurnal frequency.}\label{FFTtotal}
\end{figure*}

A Fourier transform (Fig. \ref{FFTtotal}) of the radial wind time series (day 15 to 200) in the box reveals a very strong power signal at the inertial frequency at higher latitudes in the $R_\textrm{large}$ set. Most simulations also exhibit a (much weaker) peak in power at the diurnal cycle in both domain-size sets. The $R_{small}$ simulations do not show a signal at the inertial frequency. At the mean radius and height of the box for $R_{large}(\phi>20^\circ)$ TCs, the inertial frequency (though also a function of tangential wind and its radial shear) is essentially the same as the Coriolis frequency, as indicated by the $I^2=f^2$ line which passes through the box (Figs. \ref{gstreams_longavg}). Within the cyan line, $I^2>f^2$. This line approximately follows a contour of constant tangential wind $v$ as also observed in an azimuthally averaged 3D numerical TC simulation \citep{ONeilletal2017}. Inertial waves can't propagate into a medium that has a higher inertial frequency than that of the waves, and if these waves are excited at the Coriolis frequency by the outflow at larger radii, then they would not be observed in the even higher inertial stability region near the TC core. However, even though the $I^2=f^2$ contour passes through the box of all of the smaller $R_{small}$ TCs, the outflow reaches the outer lateral sponge layer and cannot effectively excite an inertial wave.

\begin{figure*}[h]
\centerline{\includegraphics[width=29pc]{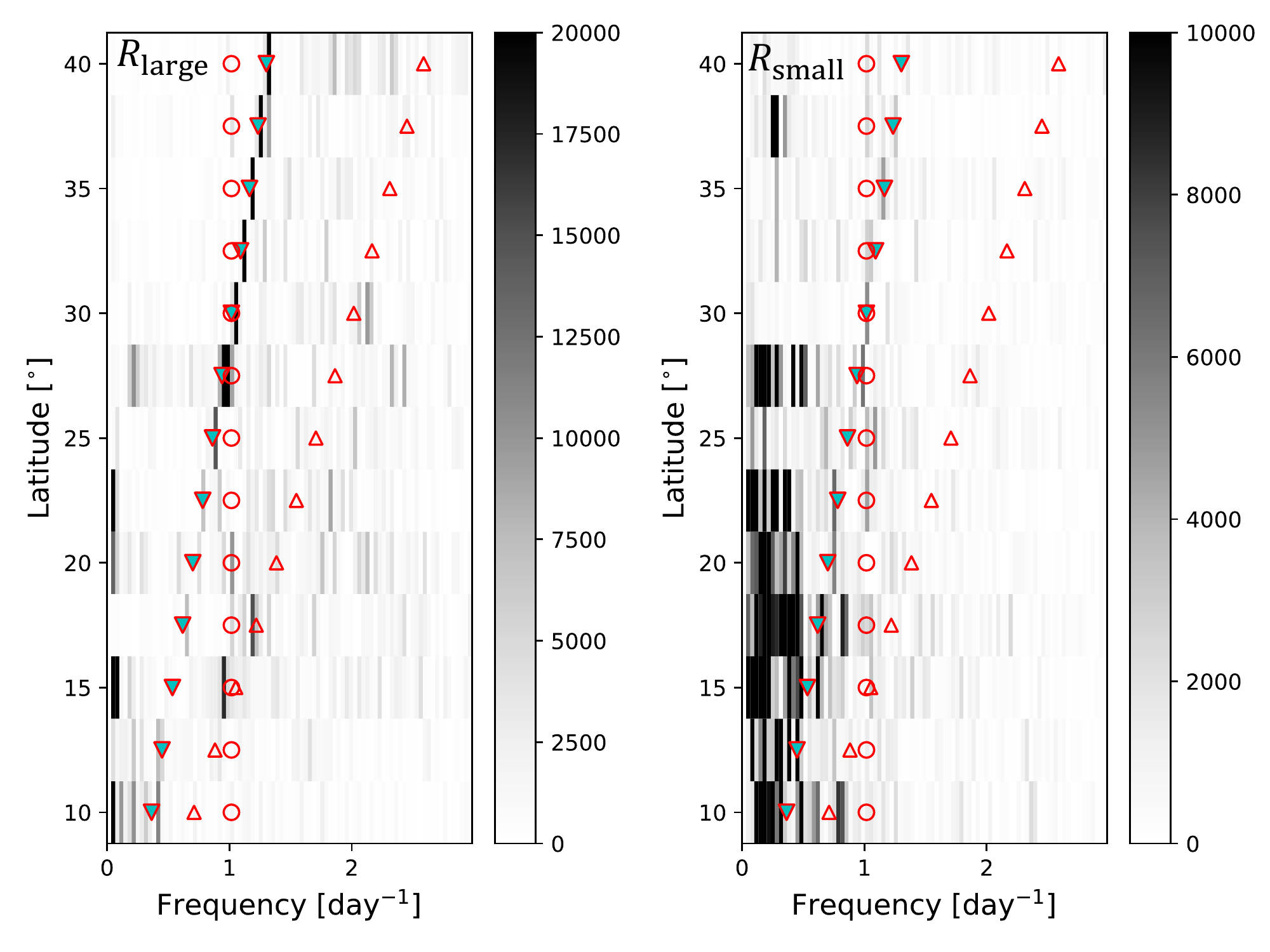}}
\caption{Power spectral density [W/Hz] of radial wind for the 30 day period of maximum inertial frequency power. Filled triangles indicate the Coriolis frequency $f$, open triangles indicate $2f$ and open circles indicate the diurnal frequency. Note the factor of two difference in colorbars.}\label{FFTinert}
\end{figure*}

Because the storms exhibit substantial structural variability over their lifetimes, we seek a period in each TC's lifetime during which the inertial wave is least affected by large-scale ERC-like events. A 30 day window that maximized power near $f^2$ was sought for each of the 26 simulations because it is long enough to allow the lowest latitude storm ($10^\circ$), with an inertial period of 2.9 days, to experience at least ten wave periods. It is also short enough to reduce the likelihood that the phase of an inertial wave is reset by an ERC-like event or other internal variability, which would reduce the magnitude of an inertial signal in a composited inertial period. The resulting power spectrum of radial wind variations in the box for the 30 day period that maximizes it is shown in Fig. \ref{FFTinert} for every simulation as calculated by a sliding 30-day window of the FFT. The identified windows for the $R_{large}$ storms are depicted in Fig. \ref{evolution} by the black boxes for latitudes $25^\circ-40^\circ$ where the inertial frequency has the highest power in the spectrum.

These 30 day periods experience some inertial wave activity at a latitude as far equatorward as 17.5$^\circ$, but the really strong wave activity is reserved for higher latitude storms. Power at the diurnal cycle is much lower than the maximized inertial frequency power. Where low latitude storms are too large to support a Coriolis-frequency wave in the box (because the wind field induces a locally higher inertial frequency), a peak at approximately $2f$ is observed. There is some power at the inertial frequency even in $R_{small}$, but it is much less intense and consistent across latitudes compared to $R_{large}$, consistent with the presence of a sponge layer to represent the environment that strongly damps wave activity. Instead, power spectra for both the long time series and the short window of $R_{small}$ simulations exhibit substantial low frequency variations, in contrast to $R_{large}$ simulations where power tends to peak at or beyond the Coriolis frequency.

\begin{figure*}[h]
\centerline{\includegraphics[width=39pc]{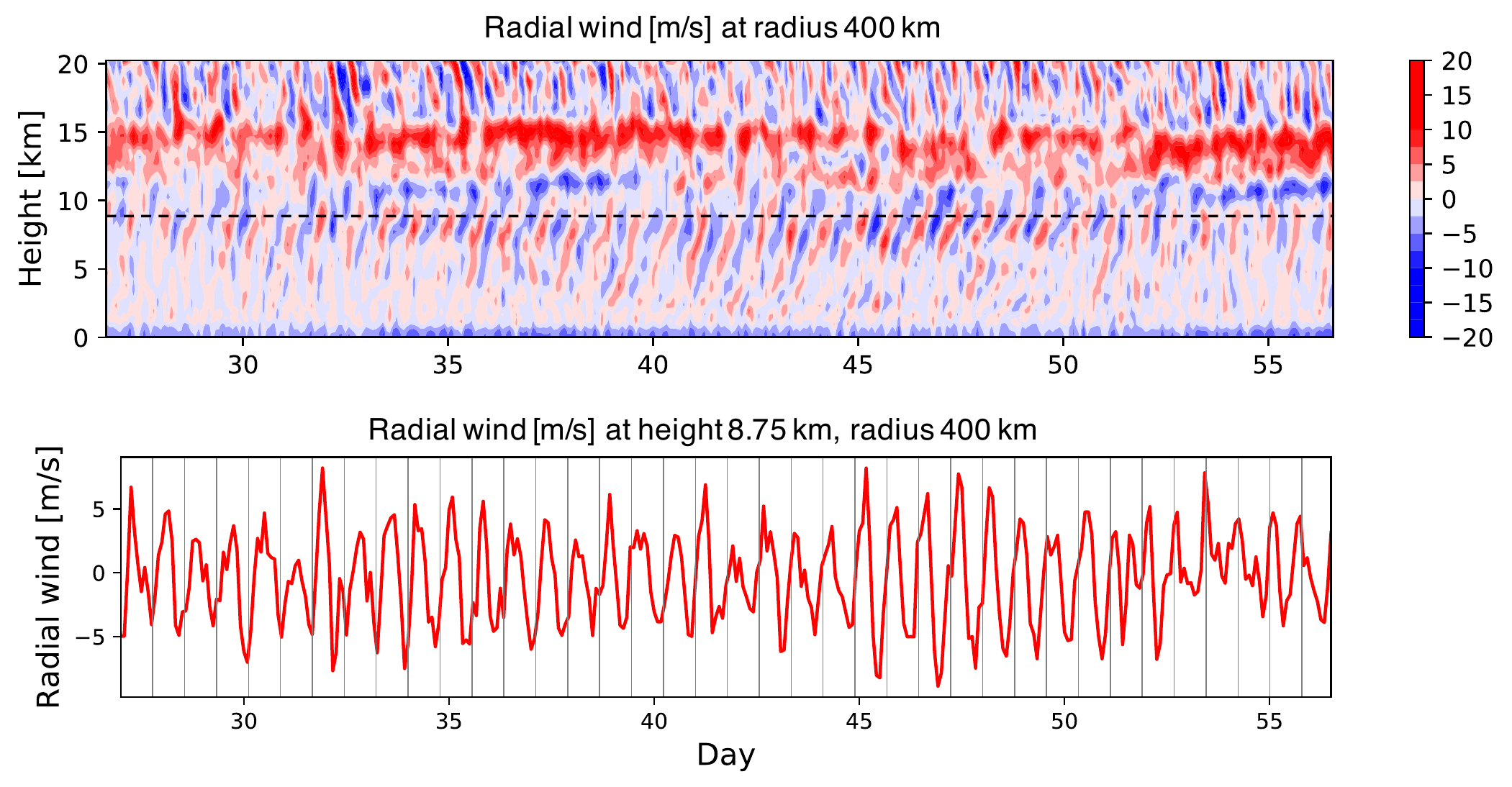}}
\caption{Top plot: Radial wind Hovmoller diagram for the 30 day period of peak inertial wave power for experiment $R_{ large}(40^\circ)$. Wind shown is averaged radially from 392 km to 408 km at each time step to smooth the output. The black dashed line indicates the height at which the bottom plot time series is measured. Bottom plot: The radial wind radially averaged over the same location but only for the altitude 8.75 km (the midpoint of the box). Grey lines are placed once every Coriolis period (18.7 hrs at $40^\circ$).}\label{g_radialwave}
\end{figure*}

The strongest inertial wave comes from the highest latitude storm $R_{large}(40^\circ)$. A time series of radial wind in the vicinity of the box (measured at a radius of 400 km) is shown in Fig. \ref{g_radialwave} for $R_{large}(40^\circ)$ from day 26 to 56 (the identified 30 day window for this TC). A signal in the radial wind propagates upward from above the boundary layer to the bottom of the outflow at almost exactly the Coriolis frequency. The propagation can be seen in the rightward tilt with height exhibited by contours of constant radial wind. The upward propagation of this inertial wave is consistent with a wavemaker at some height above it, which we propose is the time-dependent outflow jet. No filtering has been done of the wind other than a simple radial average from radius 392 km to 408 km to smooth out noise at each time step (2 hourly). The inertial wave is clearly the dominant variation in radial wind in the box vicinity with an amplitude of 4-6 m/s.

\begin{figure*}[h]
\centerline{\includegraphics[width=39pc]{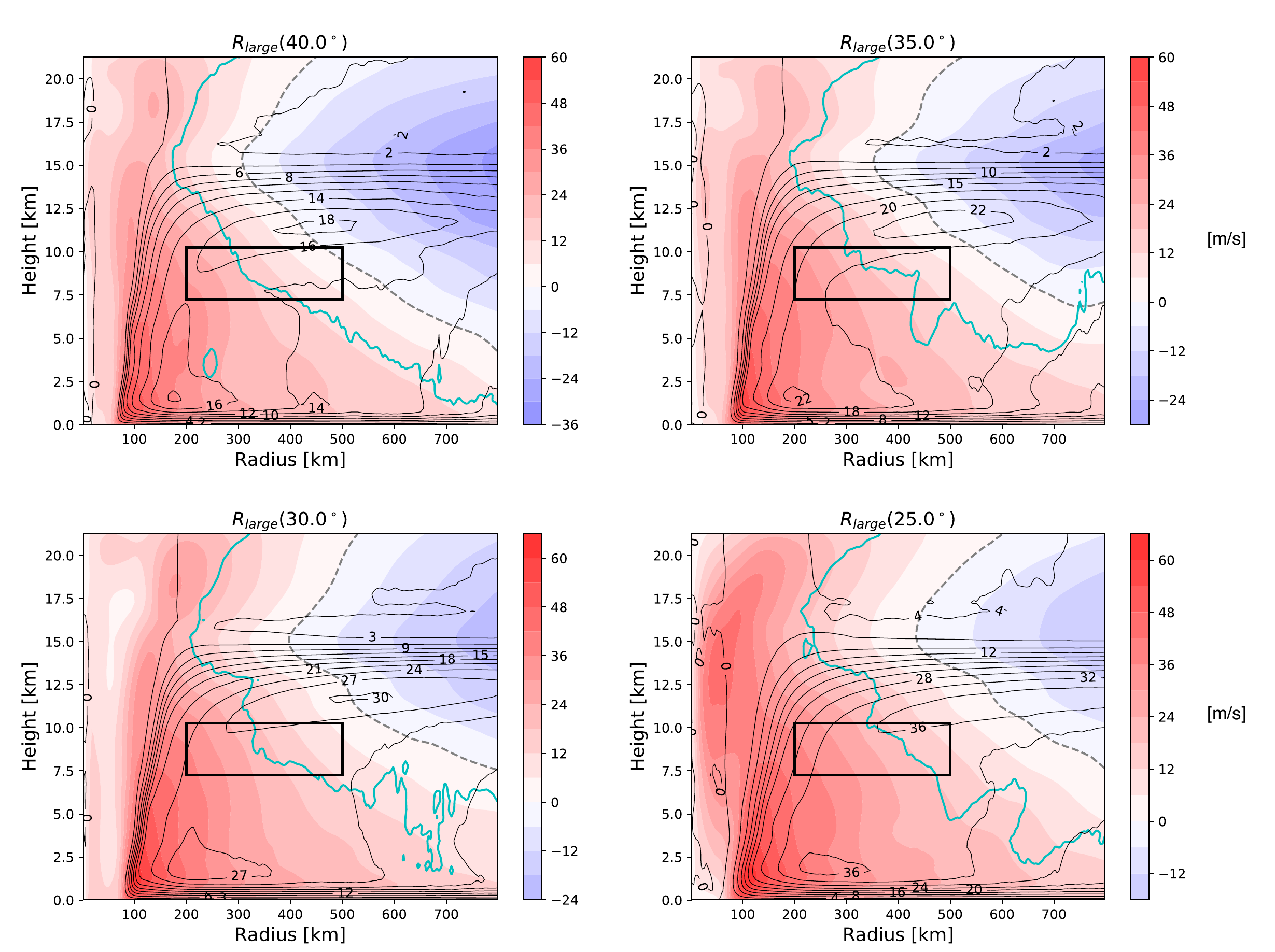}}
\caption{As Fig. \ref{gstreams_longavg} for $R_{large}$ simulations, for the 30 day period when the inertial oscillation of radial wind within the black rectangle has the highest power.}\label{gstreams_inertial}
\end{figure*}

The mean mass streamfunctions for these 30 day windows (Fig. \ref{gstreams_inertial}) indicate that the inertial wave is strongest when the TC structure is relatively steady. Though these streamfunctions depict smaller RMW and stronger winds, the time series in Fig. \ref{evolution} suggests that what we're really seeing is a reduction of variability in TC structure, and thus less smoothing of the storm in the temporal average. The inertial wave appears to be a feature of a relatively steady axisymmetric TC. Additionally the $I^2=f^2$ line has moved radially inward in each case, making it more likely that an inertial wave can be detected within the box region.

\begin{figure*}[h]
\centerline{\includegraphics[width=39pc]{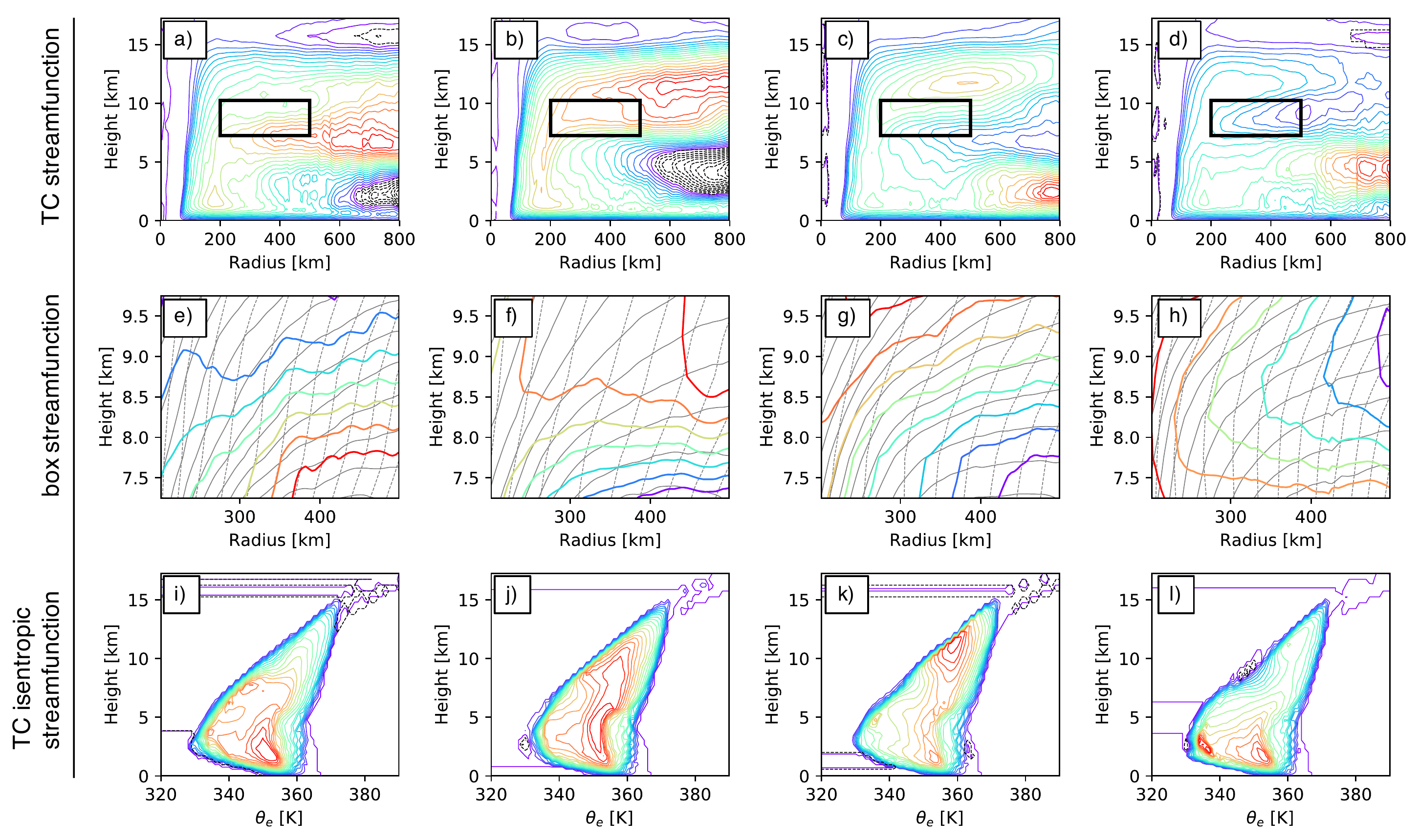}}
\caption{$R_\textrm{large}(40^\circ)$ binned into four phases (4.7 hrs apart) of the inertial period to create a composite inertial period (18.7 hrs at $40^\circ$). Top row (a-d): $(r,z)$ coordinate streamfunction. The contours occur in increments of $5\%$ of the total mass of the overturning streamfunction in the selected domain 0-800 km; negative streamfunction values are denoted by dashed black contours. Middle row (e-h): Streamfunction (thick colored lines), angular momentum surfaces (grey dashed lines), and equivalent potential temperature $\theta_e$ surfaces (grey solid lines) zoomed into the box region of the top row. Bottom row (i-l): Isentropic mass streamfunction $(\theta_e, z)$.}\label{g40_inertialstreams}
\end{figure*}

This inertial wave impacts the thermodynamic cycle of the TC. For the rest of the paper we focus on the strong wave feature in $R_{large}(40^\circ)$. We follow the procedure of isentropic analysis in \cite{PauluisMrowiec2013} and \cite{Mrowiecetal2016} to recast the TC overturning circulation into isentropic coordinates (Fig. \ref{g40_inertialstreams}). We use equivalent potential temperature with respect to water vapor (\citep{Emanuel1994book} Eq. 4.5.11)\footnote{For our purposes this is exact enough, because in the region of interest the entropy deviations with respect to ice are minimal. This is demonstrated in the next section. Therefore we don't need a more complex equivalent potential temperature with respect to ice $\theta_ei$ which includes latent heat of freezing (e.g., \citep{Pauluis2016} Eq. 2).}:

\begin{eqnarray}
\theta_e = T\frac{p_{00}}{p_d}^{R_d/C}H_{rel}^{-R_v q_v/C}\exp{\frac{L_v q_v}{CT}}
\end{eqnarray}
where $T$ is a reference temperature at the reference pressure $p_{00}=10^5$ Pa, $p_d$ is the partial pressure of dry air, $c_{pd}$, $c_{pl}$ and $c_{pi}$ are the specific heat capacities of dry air, liquid water and ice respectively, $q_v$, $q_c$ and $q_r$ are the mixing ratios of water vapor, cloud water and rain water respectively, $q_i$, $q_s$ and $q_g$ are the mixing ratios of water ice, snow and graupel respectively, $R_d$ and $R_v$ are the dry air and water vapor gas constants respectively, $C=c_{pd} + c_{pl}(q_v+q_c+q_r+q_i+q_g+q_s)$ and $H_{rel}$ is relative humidity.


Isentropic averaging is a binning of air parcels at any radius into $\theta_{e}$ buckets of some small interval $\Delta \theta_{e}$. The isentropic integral for axisymmetry is

\begin{equation}
\langle f \rangle (z,\theta_{e0},r_0,t) = 2\pi  \int \int_A f(z,r,t)\delta\{\theta_{e0}-\theta_{e}(r,z,t)\}\delta\{r_0-r\}r dr dt
\end{equation}

for any variable $f$ and Dirac delta function $\delta\{\}$. The temperature interval $\Delta\theta_{e}$ is 0.5 K and $\Delta r$ is the horizontal model resolution of 4 km. The corresponding isentropic mass streamfunction \citep{PauluisMrowiec2013,PauluisZhang2017} is

\begin{eqnarray}
\Psi(z,\theta_{e0})&=&\int_0^{\theta_{e0}} \langle \rho w \rangle(z,\theta_{e}')d\theta_{e}'\\
&=&\frac{2\pi}{TL_{outer}}\int_0^T \int_{0}^{L_{outer}}\rho w H[\theta_{e0}-\theta_{e}(r,z,t)]rdr\\
\end{eqnarray}
using the Heaviside function $H$. Instead of integrating from the outer region of the domain inward toward the center of the TC \citep{Mrowiecetal2016} we integrate from the center $r=0$ outward. For all types of streamfunctions in this paper, we calculate and show the region 0-800 km.

One challenge with the isentropic analysis procedure, particularly for large storms, is that the streamfunction in $(r,z)$ space need not actually be closed at all to achieve a closed streamfunction in $(\theta_{e},z)$ space, provided that one removes the vertical velocity averaged over the subdomain of interest \citep{Mrowiecetal2016}. Removing the mean vertical velocity closes the streamfunction regardless of whether any outflow air actually subsides within the radial limit of integration, and we do that here as well. Instead one could integrate outward all the way to the deformation radius, or where the streamfunction changes sign on average in the radial direction, but the thermodynamic signature would then be dominated by outer regular convection and the eyewall circulation would disappear.


Fig. \ref{g40_inertialstreams} shows the evolution of the composite inertial wave for $R_{large}(40^\circ)$, binned repeatedly from day 26 to 56 into four evenly spaced phases of the inertial frequency $f$ at $\phi=40^\circ$ (only nearest-neighbor snapshots were binned, leaving some output unused to be consistent with a higher resolution binning procedure later). Both $(r,z)$ space (top and middle row) and $(\theta_{e},z)$ space (bottom row) are shown.

The character of an inertial wave propagating upward from just above the boundary layer is evident. The $R_{large}(\phi_{40^\circ})$ simulation exhibits an inertial wave with a vertical wavelength that spans the depth of the troposphere. These streamfunctions represent the instantaneous phase of the inertial wave and don't represent parcel paths. Subsiding air only moves radially 50-100 km before changing direction upon wave passage, so the transient massive interior cells do not indicate a steady state recirculation.

The use of isentropic coordinates reveals that the wave does not just occur spatially but in the moist thermodynamic variable $\theta_{e}$ as well, where again its vertical wavelength spans the depth of the troposphere. A quarter of the total overturning mass circulates in just an interior upper or lower cell in $\theta_e-z$ space and about $40\%$ of the total mass in the domain appears modulated by the inertial wave. These streamfunctions look closed because of the removal of mean $w$ during the isentropic averaging process mentioned above.

The isentropic streamfunction approximates a thermodynamic diagram in $T$-$s$ space if one considers synthetic Lagrangian parcels as moving perfectly along the streamlines. This technique, MAFALDA, was developed formally by \citet{Pauluis2016} (an early approximation first appeared in \citep{Hakim2011} Fig. 9). Because temperature $T$ monotonically decreases with height in the troposphere, the bottom row of Fig. \ref{g40_inertialstreams} is effectively a $T-s$ diagram. On the other hand these phase-specific streamfunction snapshots are very transient and do not lend easy interpretation of any real parcel's trajectory because they vary much faster than air typically descends in the subsiding branch.

\begin{figure*}[h]
\centerline{\includegraphics[width=39pc]{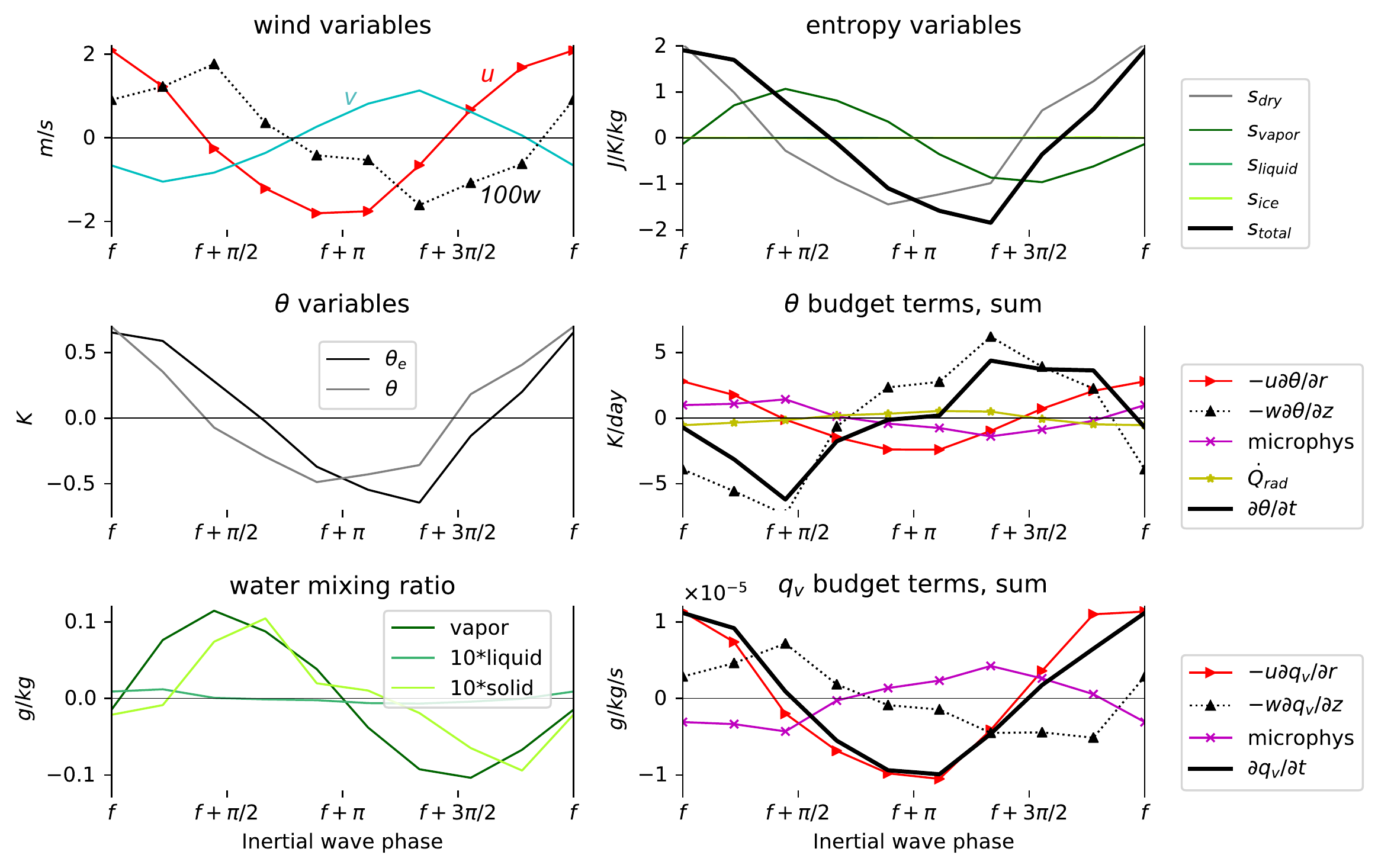}}
\caption{Departures from a composite average of an inertial period in $R_{large}(40^\circ)$ in several fields from days 26-56, binned into nine phases. The mean has been removed and the phase has been shifted such that the peak outward radial wind departure from mean is the start of the wave.  In the bottom two plots of the right column, only the non-negligible terms in the $\theta$ and $q_v$ budgets are shown; however $\partial \theta/\partial t$ and $\partial q_v/\partial t$ are the sum of every term solved by CM1 for the respective budgets.}\label{g40compositewave}
\end{figure*}

We return to a time series of mass-weighted averaged fields in the box from an Eulerian perspective. Inertial wave composites are made by binning the nearest-neighbor snapshot in time into one of nine evenly distributed phases for each inertial (Coriolis) period over 30 days. Nine bins were chosen as the highest temporal resolution possible before a particular snapshot could be binned more than once, given 2-hourly output and the approximately 19 hr inertial period at $40^\circ$. The result is a composite inertial wave cycle from which the composite average has been removed. Fig. \ref{g40compositewave} shows the wave for $R_{large}(40^\circ)$.

In the composite inertial wave, the radial wind $u$ and tangential wind $v$ are nearly in quadrature with $v$ leading by a bit more than a quarter wavelength - this is an anticyclonic oscillation in the horizontal winds, consistent with an inertial wave. There is also an oscillation in the vertical wind $w$, multiplied in Fig. \ref{g40compositewave} by 100 to be visible on the same axis. $w$ is out of phase with $v$ and in quadrature with $u$ - the latter feature being consistent with an anomalous overturning circulation as seen in Fig. \ref{g40_inertialstreams}.


The inertial wave in the box of $R_{large}(40^\circ)$ for the 30 day window occurs in the following way, arbitrarily starting at $f$ which we define as the peak radial wind outward:%

\paragraph{$f$ to $f+\pi/2$:} Radial outflow away from the storm core starts to weaken and approach zero. Horizontal advection of high-$\theta$ air into the box from the core consequently decreases. Vertical advection upward into the box from lower $\theta$ levels brings cooler air into the box and $\theta$ of the box starts to fall. However this cooling leads to condensation of water vapor and a release of latent heat (as indicated by the microphysics terms), slightly mitigating box cooling rates. Total $\theta$ tendencies reach their most negative and the box is cooling rapidly, dominated by vertical advection. Water vapor mixing ratio is still increasing temporarily because both horizontal advection from the core and vertical advection from below are fluxing moist air into the box at a higher rate than condensation is occurring. Liquid and then solid water mixing ratios increase as well as the box cools.

\paragraph{$f+\pi/2$ to $f+\pi$:} The radial wind is now increasingly negative (i.e. inflow) and starts to bring low-$\theta$ exhaust air from the outer part of the TC circulation inward. The vertical wind decreases and becomes (relatively) negative, and switches from advecting cool air from below to bringing in dry, high $\theta$ air from above. However $\theta$ in the box continues to decrease because the warming is balanced by evaporation of condensed water and corresponding latent cooling. Box-averaged $q_v$ starts to decrease as evaporation can't offset the large radial influx of very dry air from outer radii displacing moist air, and thus $\theta_e$ decreases as well.

\paragraph{$f+\pi$ to $f+3\pi/2$:} Radial inflow is strongest and now begins to weaken toward zero. $\theta$ begins to increase due to a downward advection of high $\theta$ air from above, even as evaporation continues to cool the domain, because radial advection of cool air weakens and approaches zero. Water vapor mixing ratio decreases rapidly toward its lowest level in spite of the maximum in evaporation rate in the box associated with dry outflow air being advected both downward and inward into the box.

\paragraph{$f+3\pi/2$ to $f$:} Radial wind increases from near-zero to its peak outward value. The $\theta$ perturbation becomes positive and rises to its highest value, as initially both radial advection outward and vertical advection downward bring high $\theta$ air into the box. When $w$ switches sign and starts to bring cooler air from below, $\theta$ levels off. Water vapor steadily increases back to its mean value due to a renewed horizontal advection of very moist air from the core into the box. In this phase, box evaporation followed by condensation is roughly balanced by the changing sign of the vertical advection term.\\

The wave tendency of dry potential temperature $\theta$ is always dominated by vertical advection, whereas the wave tendency of water vapor mixing ratio $q_v$ is almost entirely due to radial advection (because vertical advection typically cancels the impact from phase changes). It can also be seen that the composite wave time series of $\theta_e$ and $s$ have effectively identical behavior. The contributions to box entropy due to wave-anomalous liquid and solid phases of water are virtually zero so we can approximate total entropy with $\theta_e$ for this analysis.

\section{Summary and Discussion}\label{conclusion}

\begin{figure}[h]
\centerline{\includegraphics[width=19pc]{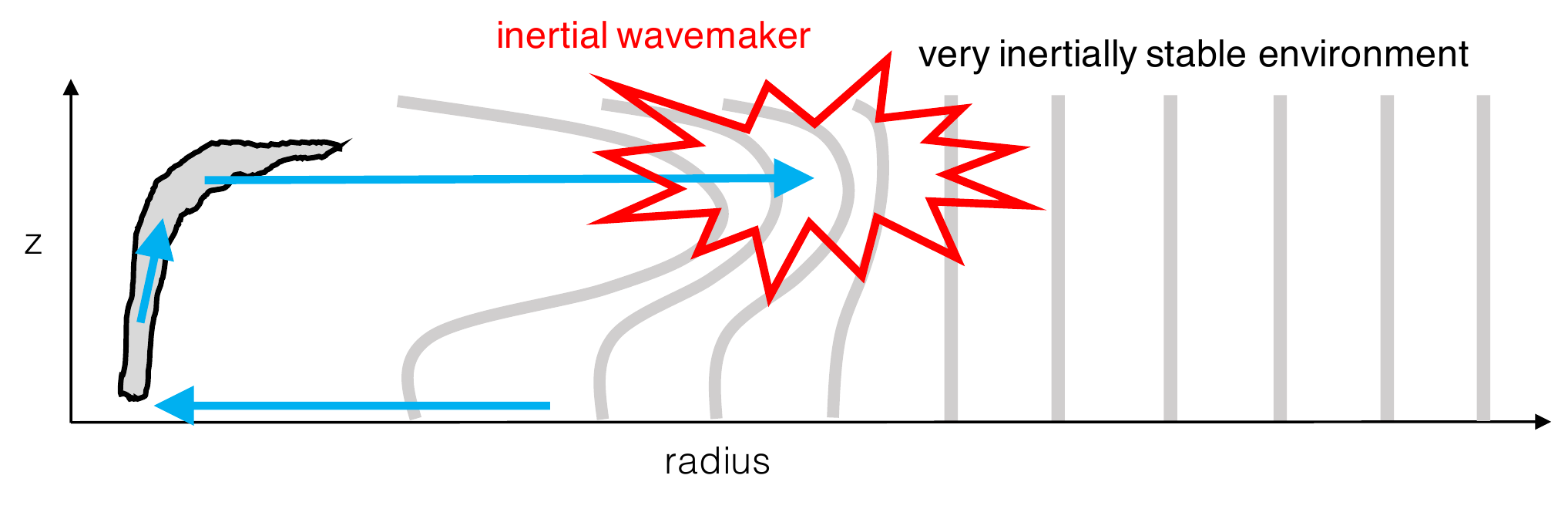}}
\caption{A schematic of the outflow wavemaker mechanism. Grey lines arre surfaces of constant angular momentum. Blue lines indicate the overturning circulation. The red burst indicates where the unsteady outflow strikes and excites the quiescent environment.}\label{wavemaker}
\end{figure}

\citet{RuppertONeill2019} found a diurnal oscillation between one and two closed cells in the overturning streamfunction of a simulated TC due to daytime heating and nighttime cooling. This was interpreted as the full-depth response to the TC canopy diurnal wave \citep{Dunionetal2014}. \citet{RuppertHohenegger2018} previously found the same diurnal overturning oscillation in a simulated nonrotating organized convective system. Like the present work, those overturning cells vary much faster than the subsidence time for outflow air to return to the boundary layer, so the overturning cells indicate a radial oscillation of subsiding air. We find here that the inertial wave at high latitudes in large axisymmetric domains can also induce a strong radial oscillation in flow outside of the TC core, at altitudes below the `wavemaker' of the outflow jet striking the inertially stable environment (Fig. \ref{wavemaker}). The inertial wave doesn't occur in smaller domains because the environment, dominated and strongly damped by the sponge layer, cannot push back on the storm in response to outflow forcing.

Our primary findings can be summarized as following:
\begin{itemize}
\item Simulated TCs at higher latitudes experience an inertial wave excited by the collision of the outflowing air with the highly inertially stable environment;
\item The inertial wave dominates structural periodicity in spite of an interactive diurnal cycle during times of relatively little secular TC variation;
\item The inertial wave is a mechanism of TC self-ventilation \citep{TangEmanuel2010} in the absence of any environmental vertical wind shear;
\item This self-ventilation periodically decreases entropy near the core of the storm at midlevels; and
\item The wave does not appear to impact streamlines experiencing the most significant temperature difference in the first place: those in the core of the eyewall. Thus, the thermodynamic circuits associated with the strongest inner-core winds (the circuits which constitute a Carnot-cycle constraint on the maximum tangential surface winds \citet{BisterEmanuel1998,Hakim2011,RousseauRizziEmanuel2019}) remain unchanged.
\end{itemize}

These findings are robust across mid-high latitude simulations, and evident as low as $22^\circ$ latitude, if the domain extent exceeds the Rossby deformation radius. There are weak indications that this wave occurs in small domains too but the physical mechanism that makes it possible - a resonance with the inertially stable environment - is nearly prohibited in such models.

We also looked at whether the inertial wave propagated throughout the environment by examining the radial wind FFT of a similarly sized box centered at 4200 km (and same height; not shown), well beyond most deformation radii and all $I^2=f^2$ contours. We found that it did at every latitude, albeit with lower power distributed more broadly around the Coriolis frequency. The highest and most concentrated power remains at the highest latitudes. This is consistent with our finding that the wave is prohibited from regions where $I^2>f^2$, and demonstrates that even low-latitude axisymmetric storms can excite an inertial wave beyond where Coriolis-frequency oscillations are not precluded by cyclonic winds. However, we study in this paper a box that is limited to 500 km in radius because the wave is reliant on conservation of angular momentum and therefore a substantial degree of axisymmetry. This allows some potential for comparison of these results with 3D TCs, both observed and simulated. Because the axisymmetric high latitude TCs experience the inertial wave in a region where axisymmetry is not too severe an assumption (within 500km radially), a highly inertially stable environment appears the most likely place to find such behavior in nature.

The wave period is much shorter than the time it takes for a parcel of air to subside from the outflow back to the boundary layer (two-three weeks as estimated by parcel tracking). Every simulation in this study included 500 parcels allowed to freely circulate within the TC over the course of the 200 day integration. A Lagrangian analysis of the parcels themselves in contrast with the synthetic parcel paths of the MAFALDA procedure will be the subject of future work.

Numerical simulations of single, axisymmetric TCs often occur in domains constrained by computing expense, with open or closed boundaries placed 1000-1500 km from the storm center (e.g., \citet{RotunnoEmanuel1987,BryanRotunno2009,Smithetal2011,Hakim2011,EmanuelRotunno2011,Hakim2013,Ramsay2013,NavarroHakim2016,Allandetal2017,RousseauRizziEmanuel2019}). 3D simulations are necessarily even smaller - commonly run in ~2000x2000 km$^2$ doubly periodic domains, or half the radial extent of most axisymmetric runs. Yet the default simulated latitude for these storms is $20^\circ$, and the deformation radius in the quiescent tropical atmosphere at $20^\circ$ is approximately 2,500 km -- double the extent allowed for the secondary circulation of these simulated storms. A limited number of recent studies has used much larger domains for idealized but well-resolved TCs, including \citep{Rappinetal2011,Nolan2011,ChavasEmanuel2014,Wangetal2014,Frisius2015,PauluisZhang2017,Daietal2017,Persingetal2019}. For 3D simulations, using a domain smaller than the deformation radius implies that the outflow will interact with itself through the periodic boundary conditions, though it is unclear whether this has any implications for studying boundary layer and inner core behavior.

Achieving a truly steady state TC over long periods of time without squeezing it into a too-small domain appears difficult when interactive radiation is employed. \citep{Hakim2011} concluded that the tropical atmosphere is unstable to axisymmetric TCs. \citet{Hakim2013} simulated TCs in a 1500 km-radial axisymmetric domain with RRTM radiation and observed regular (but aperiodic) ERCs occurring with a frequency around 4-8 days. Here, in a domain of the same radial extent, $R_{small}$ ERC-like events disrupt the TC with a recovery of up to a few tens of days. Once the TCs are given more space than they can fill, however, TCs can take 50 days or more to recover a similar maximum tangential wind speed and RMW as before the ERC-like event occurred.

Apart from the finding of an inertial wave, we have shown that the outflow size of an axisymmetric TC scales with the deformation radius. \cite{ChavasEmanuel2014} studied the impact of the Coriolis parameter on boundary layer outer winds and also found a strong $1/f$ dependence; however the full scaling was $v_p/f$, including a potential intensity $v_p$ in the numerator, such that $L_D$ was not the principal scale for the outer wind field in the boundary layer. We find that in massive domains ($R>L_D$), storm size as defined by the location of the peak anticyclonic outflow jet follows the deformation radius. This yields the possibility that the surface and upper circulations do not scale together, which deserves further study.  This may be relevant to 3D rotating radiative convective equilibrium simulations where multiple TCs fill the domain yet inter-storm spacing does not seem to scale neatly with either $v_p/f$ or $L_D$ \citep{Zhouetal2014,CroninChavas2019}. Given that the gravity wave phase speed in $L_D$ is generally larger than $v_p$ by roughly a factor of two, these findings suggest that adjacent storms would most likely interact via their outflows.

Real TC outflow is fundamentally asymmetric. As a first step toward identifying the impact of inertial stability on the outflow, we focus on surprising dynamical responses to high-latitude environments. To what extent does this inertial wave have any bearing in three dimensions? Does it occur in some three-dimensional analog form? Do outflowing jets support any kind of inertial stiffness that could provide a restoring force for an inertial wave? The similarities between the findings here and the diurnal overturning oscillation in \citep{RuppertONeill2019} suggest that 3D storms may also be impacted by substantial, periodic reversals in radial flow in the subsidence region, and future work will explore the modification and relevance of inertial waves in a highly asymmetric environment.

Returning to the \citet{Rappinetal2011} claim that subsidence is forced in strongly inertially stable environments, we don't find evidence for it. Storm outflow size robustly scales with $L_D$ and outflow remains strong throughout the integration period, whereas forced subsidence would eventually weaken and shrink TCs as the outer buoyancy profile increases to rival the buoyancy of the core. Previous work has shown that actual subsidence velocity compares well with that predicted from simple radiative-subsidence balance \cite{ChavasEmanuel2014,Chavasetal2015,ReedChavas2015,Davis2015,CroninChavas2019}Also, even our highest inertial stability values at $40^\circ$ are much lower than $N^2$, indicating that radial motions are a less expensive, more likely response than forced subsidence of buoyant air. But the claim that such environments do induce some total work cost to the TC seems correct. However since the impacted portion of the streamfunction does not occur in the extremal part of the overturning circulation associated energetically with the strongest winds \citep{BisterEmanuel2002}, i.e. traversing through the bottom of the boundary layer, into the eyewall, and out at the top of the outflow, it does not exhibit an impact on peak tangential wind speed or the RMW.



%


\acknowledgments
The authors are grateful to George Bryan for freely providing the CM1 model code, and for insightful conversations and advice from Olivier Pauluis, Leif Thomas and Malte Jansen. The authors declare no conflict of interest. MON completed part of this work while funded by the T. C. Chamberlin Postdoctoral Fellowship in the Department of the Geophysical Sciences at the University of Chicago.



 \appendix[A] 

\appendixtitle{Entropy definition}

Following \citet{Pauluis2016} (Appendix) we use as the reference state liquid water at the freezing temperature, and define specific entropy as:

\begin{equation}
s = s_d + r_v s_v + r_l s_l + r_i s_i,
\end{equation}

the sum of specific dry entropy, specific entropy of water vapor, specific entropy of liquid water and specific entropy of ice, respectively. Those entropies are defined as

\begin{eqnarray}
s_d =  c_p \ln{\frac{T}{T_f}}
s_v = c_{pl} \ln{\frac{T}{T_f}} + \frac{L_v}{T} - R_v \ln H_{rel}\\
s_r = c_{pl} \ln{\frac{T}{T_f}} \\
s_i = c_{pi} \ln{\frac{T}{T_f}} - \frac{L_{f0}}{T_f}.
\end{eqnarray}

for specific heats of dry air $c_p$, liquid water $c_{pl}$, frozen water $c_{pi}$, latent heat of vaporization $L_v$, latent heat of freezing $L_{f0}$, gas constant $R_v$, relative humidity $H_{rel}$, and the freezing temperature $T_f=273.15$K.

%
%
%


 \bibliographystyle{ametsoc2014}

 \bibliography{}

%
%

\end{document}